\documentclass[10pt]{article} 
\usepackage{amssymb}
\usepackage{amsmath}
\usepackage{amsthm}
\usepackage{latexsym} 
\usepackage[dvips]{epsfig}
\usepackage{mathrsfs}
\usepackage{eufrak}

\theoremstyle{plain}
 
\newtheorem{lemma}{Lemma}
\newtheorem{theorem}{Theorem}

\newtheorem*{conjecture}{Conjecture}

\font\SYM=msbm10 
\newcommand{\Real}{\mbox{\SYM R}}
\newcommand{\Complex}{\mbox{\SYM C}}

\font\tenscr=rsfs10 scaled1100
\font\sevenscr=rsfs7 
\font\fivescr=rsfs5 
\skewchar\tenscr='177 
\skewchar\sevenscr='177
\skewchar\fivescr='177 
\newfam\scrfam 
\textfont\scrfam=\tenscr
\scriptfont\scrfam=\sevenscr 
\scriptscriptfont\scrfam=\fivescr

\def\scri{{\fam\scrfam I}}

\def\wth{\widetilde{h}}
\def\wtD{\widetilde{D}}
\def\wtK{\widetilde{K}}
\def\wtchi{\widetilde{\chi}}

\def\O{\mathcal{O}}

\begin{document}

\title{\textbf{On smoothness-asymmetric null infinities}}

\author{Juan Antonio Valiente Kroon \thanks{E-mail address:
 {\tt j.a.valiente-kroon@qmul.ac.uk}} \\
School of Mathematical Sciences,\\ Queen Mary, University of London,\\
Mile End Road,\\ 
London E1 4NS,
\\United Kingdom.}

\maketitle

\begin{abstract}
  We discuss the existence of asymptotically Euclidean initial data
  sets to the vacuum Einstein field equations which would give rise
  (modulo an existence result for the evolution equations near spatial
  infinity) to developments with a past and a future null infinity of
  different smoothness. For simplicity, the analysis is restricted to
  the class of conformally flat, axially symmetric initial data
  sets. It is shown how the free parameters in the second fundamental
  form of the data can be used to satisfy certain obstructions to the
  smoothness of null infinity. The resulting initial data sets could
  be interpreted as those of some sort of (non-linearly) distorted
  Schwarzschild black hole. Its developments would be so that they
  admit a peeling future null infinity, but at the same time have a
  polyhomogeneous (non-peeling) past null infinity.
\end{abstract}


\section{Introduction}
This article is concerned with providing an example of
asymptotically Euclidean, conformally flat initial data sets for the
Einstein vacuum equations which are expected to have time developments
for which the two disconnected parts of null infinity have different
degrees of smoothness.

The analysis of the behaviour of the gravitational field in the region
of  spacetime near spatial infinity and null infinity carried out in
\cite{Fri98a, Val04a, Val04d} ---for spacetimes with time reflexion
symmetry--- and in \cite{Val04e, Val05a} ---for spacetimes without
time  reflexion--- was motivated by the desire of setting on a sound
footing  \emph{Penrose's proposal} for the description of the
asymptotics of the  gravitational field ---see e.g. \cite{Pen63,
Fri03a, Fri04}. Penrose's proposal  suggests that the asymptotic
gravitational field of isolated systems should admit a \emph{smooth
conformal completion} at null infinity.

A large class  of spacetimes satisfying Penrose's proposal has been
constructed by Chru\'{s}ciel  \& Delay  \cite{ChrDel02}. The examples
provided rely in a refinement of a  gluing construction introduced by
Corvino \cite{Cor00} by means of which it  is possible to construct
time symmetric initial data sets which are  essentially arbitrary
inside a compact set, but are exactly Schwarzschild  in the
asymptotic region. More recently, a further development in the gluing
techniques has allowed to extend Corvino's results to the non-time
symmetric  case, so that initial data sets which are arbitrary in a
compact region can  be glued to a stationary asymptotic region
\cite{ChrDel03,CorSch03}. As a result, all of the available examples of 
spacetimes satisfying Penrose's proposal are stationary in a spacetime
neighbourhood of spatial infinity. In view of this, it is natural to
ask whether initial data sets with stationary asymptotic regions are
the only ones giving rise to spacetimes with a smooth null
infinity. In relation to this latter issue, evidence for the following
conjecture has been provided \cite{Val04a,Val04e}:

\begin{conjecture}
The time development of an asymptotically Euclidean initial data set
which is conformally flat in a neighbourhood of infinity admits a
conformal extension to both future and past null infinity of class
$C^k$, with $k$ a non-negative integer, if and only if the initial
data are Schwarzschild to order $p_*$ where $p_*=p_*(k)$ is a
non-negative integer. If the development admits an extension of class
$C^\infty$, then the initial data have to be exactly Schwarzschild on
$\mathcal{B}_a(i)$.
\end{conjecture}

Here, for \emph{Schwarzschild up to order $p_*$} it is understood
that asymptotic expansions of the initial data near infinity coincide
up to order $\mathcal{O}(1/|y|^{p_*})$ with those of initial data for
the Schwarzschild solution ---see section 2 for more details. A
similar conjecture is expected to hold for non-conformally flat
initial data ---so that the data should be stationary up to order
$p_*$ if a null infinity of class $C^k$ is to be attained.

The calculations leading to the conjecture have shown the possibility
of having a spacetime where future null infinity ($\scri^+$) and past
null infinity ($\scri^-$) have different smoothness ---note that the
conjecture requires conditions on both parts of null infinity, in
stark contrast with a previous version which arose in the analysis of
time symmetric situations \cite{Val04a}. The possibility of having a
spacetime where the two components of null infinity have different
degrees of differentiability arose for the first time in the
post-Newtonian analysis of the relativistic Kepler problem carried out
by Walker \& Will \cite{WalWil79a, WalWil79b}. Their calculations
exhibit a system whose fall-off at null infinity is compatible with
the \emph{Peeling Behaviour} at $\scri^+$, but not at $\scri^-$. The
latter phenomenon was regarded as a hint that requiring smoothness of
the gravitational field at past null infinity may be a too stringent
condition ---see e.g. \cite{PenRin86}.  For many physical applications
the presence of a non-smooth $\scri^-$ may not be too hampering as
long as one can ensure a peeling $\scri^+$. And even if $\scri^+$
happens to be non-smooth, most of the relevant structure at null
infinity can still be recovered ---see \cite{ChrMacSin95}. As pointed
out by Penrose: ``the issue is not whether regularity at $\scri^+$
covers all situations we would like to call \emph{asymptotically
flat}; it is whether this regularity condition allows all the freedom
that we need in order to describe isolated systems in general
relativity'' ---see \cite{Pen02}.

In \cite{Val04e}, scripts in the computer algebra system {\tt Maple V}
were used to calculate asymptotic expansions of the time development
of conformally flat (but non-time symmetric) initial data sets in a
neighbourhood of spatial infinity. In particular, the components
$\phi_j$ with $j=0,\ldots,4$, of the Weyl spinor $\phi_{ABCD}$ have an
expansion of the form:
\[
\phi_j \sim \sum_{p=|2-j|}^\infty \phi^{(p)}_j \rho^p,
\]
where $\rho$ is a geodesic distance, and the coefficients
$\phi^{(p)}_k$ depend on $\tau$, an affine parameter of conformal
geodesics, and on some angular coordinates. In these coordinates
$\tau=\pm 1$ corresponds to the locus of $\scri^+$ and $\scri^-$,
respectively. It turns out that for $p=0,\ldots,4$, the coefficients
$\phi^{(p)}_j$ extend smoothly through null infinity. However, from
$p=5$ onwards, one finds a series of \emph{obstructions} to the
smoothness of both future and past null infinity. For $p=5$ the
obstructions are time symmetric ---that is, the obstructions for
$\scri^+$ and $\scri^-$ are the same, but at $p=6$ some of the
obstructions do not possess this time symmetry. The latter ---as it
was discussed in previous paragraphs--- opened the possibility of
having a future and a past null infinity with different degrees of
smoothness. A similar phenomenon occurs, as it is to be expected, if
one analyses time asymmetric initial data sets which are not conformally
flat. However, the overall picture is much more involved.

The calculations in \cite{Val04e, Val05a} assume the existence of
initial data sets which are expandable in powers of $1/r$ near
infinity. Existence proofs for this type of data have been provided in
\cite{DaiFri01,Dai04a}. However, the question whether the free
parameters in the solution can be chosen such that the conditions
leading to smoothness-asymmetric null infinities are satisfied
remained open. This article provides an answer to this
question. Ultimately, one would also like to make some assertion
concerning the existence of the time development. This question is a
much more difficult one and lies outside the scope of the present work.

\bigskip
The article is structured as follows. Section 2 is concerned with some
aspects of conformally flat initial data sets which will be required
in our investigations. In particular, some solutions to the momentum
constraint more general than those of the Bowen-York Ansatz are
considered, and a general existence result for the class of initial data
under consideration is recalled. In section 3 we discuss a certain class
of obstructions to the smoothness of null infinity which allow us to
construct initial data sets whose developments would have prescribed
regularity at null infinity. Section 4 contains our main result and
its proof.

The present work is a natural extension of the work carried out  in
\cite{Val04e,Val05a}. We have endeavour to follow to the nomenclature and
notation of these references, however, for convenience we have made
use of spin-weighted harmonics instead of certain unitary
representations of $SU(2,\Complex)$ when performing
expansions. Accordingly, our notation has been adapted. The reader in
need of a remainder on concepts and results on functional analysis
(Sobolev spaces, G\^ateaux derivatives, the implicit function theorem)
is remitted to \cite{AmbPro95,Aub82,GilTru98}.

\section{Framework}

We shall restrict our discussion to 
solutions
$(\widetilde{h}_{ab},\widetilde{\chi}_{ab})$ to the
vacuum Einstein constraint equations:
\begin{eqnarray}
&&\wtD^a \wtchi_{ab}-\wtD_b \wtK=0 \label{momentum_constraint}, \\
&&\widetilde{r} - \wtK^2+\wtK_{ab}\wtK^{ab}=0 \label{hamiltonian_constraint},  
\end{eqnarray}
on asymptotically Euclidean 3-manifolds, $\widetilde{\mathcal{S}}$,
which are \emph{maximal} ---that is, $\wtK=0$--- and \emph{conformally
flat}. The metric $\widetilde{h}_{ab}$ will be taken to be \emph{negative
definite}. For simplicity, we shall also assume that the initial data
sets are axially symmetric. Nevertheless, it must be pointed out that
our analysis can be generalised to the axially symmetric and/or
conformally flat settings. Furthermore, we shall assume that the
3-manifold $\widetilde{\mathcal{S}}$ has the topology of the time
symmetric slice of the Schwarzschild solution: two asymptotically flat
regions connected by a ``throat ---see figure 1''.

\begin{figure}[t]
\centering
\includegraphics[width=0.3\textwidth]{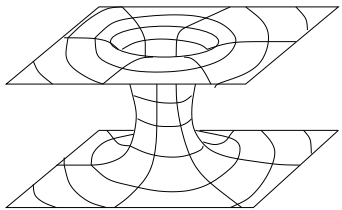} 
\caption{The topology of the initial hypersurface $\widetilde{\mathcal{S}}$ corresponds to that of time symmetric slices of Schwarzschild: two asymptotic regions connected by a ``throat''.} 
\end{figure}

The construction we want to
implement depends crucially on the properties of the solutions to the
constraints in a neighbourhood of infinity. Therefore, we adopt the
conformal compactification  picture. Let $\mathcal{S}$ be a smooth, 
orientable, connected, compact 3-manifold ---for definiteness, we can think 
of $S^3$. In $\mathcal{S}$ we shall consider two special points $i$ 
and $i^\prime$ ---say, the North and South poles of $S^3$--- which will 
represent the two infinities of the asymptotic ends of the physical 
data $(\widetilde{\mathcal{S}},\wth_{ab},\wtchi_{ab})$, and so that
$\widetilde{S}=S\setminus \{i,i^\prime\}$.

\subsection{The conformal method}
Following the standard approach ---\emph{the Conformal Method}--- we set
\begin{equation}
\wth_{ab}=\vartheta^4 h_{ab}, \quad \wtchi_{ab}=\vartheta^{-2}\psi_{ab}.
\end{equation}
Let $x^a$ denote $h$-normal coordinates centred at $i$. When no
confusion arises, we may write $x$ instead of $x^a$. In order to
ensure the asymptotic Euclideanity of the initial data we require
\begin{subequations}
\begin{eqnarray}
&& \psi_{ab}=\mathcal{O}(|x|^{-4}) \mbox{ as } |x|\rightarrow 0, \label{boundary_1} \\
&& \lim_{|x|\rightarrow 0} |x|\vartheta = 1, \label{boundary_2}
\end{eqnarray}
\end{subequations} 
with $|x|^2=\delta_{ab}x^a x^b$. Similarly, if $y^a$ denote $h$-normal
coordinates centred at $i^\prime$, we shall require that
\begin{subequations}
\begin{eqnarray}
&& \psi_{ab}=\mathcal{O}(|y|^{-4}) \mbox{ as } |y|\rightarrow 0, \label{boundary_3} \\
&& \lim_{|y|\rightarrow 0} |y|\vartheta = c^\prime, \label{boundary_4}
\end{eqnarray}
\end{subequations} 
where $c^\prime$ is a positive constant.

Because of the conformal flatness of the data, we set in a
neighbourhood of $i$, $h_{ab}=-\delta_{ab}$, so that near $i$ the
constraint equations reduce to:
\begin{subequations}
\begin{eqnarray}
&& \partial_a \psi^{ab}=0, \label{cf_momentum}\\
&& \Delta \vartheta=-\frac{1}{8}\psi_{ab}\psi^{ab}\vartheta^{-7}, \label{cf_hamiltonian}
\end{eqnarray}
\end{subequations}
where $\Delta$ denotes the flat Laplacian.

\subsection{Solutions to the momentum constraint}
The solutions to the momentum constraint in Euclidean space have been 
extensively studied in the literature ---see e.g. 
\cite{DaiFri01,BeiOMu96,Val04e}. For convenience, we will use 
the version given in \cite{Val04e}, in which the solutions to the momentum 
constraint are derived using the space-spinor formalism 
---see \cite{Som80,Fri98a}. For this, 
we consider the vector field $e_{(3)}^a$ which 
in Cartesian coordinates is given by $x^a/|x|$, and complete it to an 
orthonormal basis $\{e_{(1)}^a, e_{(2)}^a, e_{(3)}^a\}$. Let 
$\psi_{(i)(j)}=\psi_{ab}e^a_{(i)}e^b_{(j)}$ denote the components of the 
second fundamental form $\psi_{ab}$ with respect to this basis. In a 
space-spinor formalism, and under the assumption of a maximal slice, the 
tensor $\psi_{ab}$ will be represented by means of 
a totally symmetric spinor $\psi_{ABCD}$ which is related to $\psi_{(i)(j)}$ 
via the spatial Infeld-van der Waerden symbols $\sigma_{(i)}^{AB}$:
\begin{equation}
\psi_{(i)(j)}=\sigma_{(i)}^{AB} \sigma_{(j)}^{CD}\psi_{ABCD}.
\end{equation}
The spinor $\psi_{ABCD}$ being totally symmetric can be decomposed as
\begin{equation}
\psi_{ABCD}=\psi_0\epsilon^0_{ABCD}+\psi_1\epsilon^1_{ABCD}+\psi_2\epsilon^2_{ABCD}+\psi_3\epsilon^3_{ABCD}+\psi_4\epsilon^4_{ABCD},
\end{equation}
where the spinors $\epsilon^i_{ABCD}$, $i=0,\ldots,4$ are also totally 
symmetric and 
\begin{equation}
\epsilon^k_{(ABCD)_j}=\delta^k_j{\binom{4}{k}}^{-1},
\end{equation}
the subindex $j=0,\ldots,4$ in ${}_{(ABCD)_j}$ indicating that $j$ of 
the indices have to be set equal to $1$. The particular solution to 
the momentum constraint that we are going to consider can be split in 
six independent pieces:
\begin{equation}
\psi_{ABCD}=\psi^\mathcal{A}_{ABCD}+\psi^\mathcal{J}_{ABCD}+\psi^{-2,2}_{ABCD}+\psi^{-1,2}_{ABCD}+\psi^{-1,3}_{ABCD}+\psi^{0,2}_{ABCD}. \label{psi_ab}
\end{equation}
The term $\psi^\mathcal{A}_{ABCD}$ associated with the expansion
conformal Killing vector ---it gives rise, for example, to time
asymmetric slices in Schwarzschild
\cite{BeiOMu98,EstWahlChrDeWSmaTsi73,Rei73}. The term $\psi^\mathcal{J}_{ABCD}$
encodes the angular momentum content of the data. While the remaining
terms are diverse multipolar terms ---the first superindex denotes the
decay of the term ($|x|^{-2}$, $|x|^{-1}$ or $|x|^0$) and the second
its multipolar nature: quadrupolar ($2^2$) or octupolar ($2^3$). For
further discussion see \cite{Val04a}.

In the case of the $\psi^\mathcal{A}_{ABCD}$ term, one has a single
non-vanishing component:
\begin{equation}
\psi^\mathcal{A}_2=-\frac{\mathcal{A}}{|x|^3}.
\end{equation}
Assuming axial symmetry, the non-vanishing components of
$\psi^\mathcal{J}_{ABCD}$ are given by:
\begin{equation}
\psi_1^\mathcal{J}=\frac{12}{|x|}\sqrt{\frac{\pi}{5}}\mbox{i} \mathcal{J}\; {}_{1}Y_{20}, \quad \psi_3^\mathcal{J}=-\frac{12}{|x|}\sqrt{\frac{\pi}{5}}\mbox{i} \mathcal{J}\; {}_{-1}Y_{20}.
\end{equation}
For the remaining terms one has that:
\begin{equation}
\psi^{n,q}_j=(-\mbox{i})^{j+6-2q}\sqrt{\frac{4\pi}{2q+1}}L_{j,n;q}\;\;{{}_{2-j}Y_{q0}}\rho^n,
\end{equation}
for $j=0,\ldots,4$. The coefficients $L_{0,n;q}$ and $L_{4,n;q}$ are freely 
specifiable complex numbers satisfying the reality condition
\begin{equation}
L_{0,n;q}=\overline{L}_{4,n;q}. \label{reality}
\end{equation}
The remaining coefficients are calculated via the formulae:
\begin{subequations}
\begin{eqnarray}
&& L_{1,n;q}=\frac{\big( (L_{4,n;q}-L_{0,n;2q})(q+1)q+4(n+7)^2L_{0,n;q}\big)}{(n+7)(2(n+8)(n+6)-(q+2)(q-1))}\sqrt{(q+2)(q-1)}, \\
&& L_{2,n;q}=\frac{3(L_{0,n;q}+L_{4,n;q})}{2(n+7)^2-q(q+1)}\sqrt{(q+2)(q+1)q(q-1)},\\
&&  L_{3,n;q}=\frac{\big( (L_{0,n;q}-L_{4,n;2q})(q+1)q+4(n+7)^2L_{4,n;q}\big)}{(n+7)(2(n+8)(n+6)-(q+2)(q-1))}\sqrt{(q+2)(q-1)}.
\end{eqnarray}
\end{subequations}
In addition, the following \emph{regularity conditions} will be required 
to hold:
\begin{equation}
L_{0,-2;2}+L_{4,-2;2}=0, \quad L_{0,-1;3}+L_{4,-1;3}=0. \label{regularity}
\end{equation}
The latter, together with (\ref{reality}) that $L_{0,-2;2}$,
$L_{4,-2;2}$, $L_{0,-1;3}$ and $L_{4,-1;3}$ are pure imaginary. The
conditions (\ref{regularity}) are intended to preclude the appearance
of a certain type of logarithmic divergences at the sets where null
infinity touches spatial infinity ---see \cite{Fri98a,Val04e}, and
also the discussion in section 3.

From the point of view of the normal coordinates $y^a$ centred on
$i^\prime$, the different terms of $\psi_{ABCD}$ have an analogous
decay to the one near $i$:
\begin{equation}
\psi^\mathcal{A}_{ABCD}=\O(|y|^{-3}), \quad \psi^\mathcal{J}_{ABCD}=\O(|y|^{-3}), \quad
\psi^{n,k}_{ABCD}=\O(|y|^{n}).
\end{equation}
Thus, all the terms ---save for $\psi^{0,2}_{ABCD}$--- are singular at
both $i$ and $i^\prime$ ---see \cite{DaiFri01}.

For latter use, it is noted that if
$\psi_{ab}$ is of the form given by equation (\ref{psi_ab}) then:
\begin{equation}
\psi_{ab}\psi^{ab}=\psi_{ABCD}\psi^{ABCD}=\frac{1}{|x|^6}\left(\frac{\mathcal{A}^2}{6}-12\mathcal{J}^2+24\sqrt{\frac{2\pi}{5}}\mathcal{J}^2Y_{20}\right)+\O\left(\frac{1}{|x|^5}\right). \label{psipsi}
\end{equation}
Note that the above expansion truncates at order $\O(1/|x|^2)$, the
first contributions of the higher multipolar terms in
$\psi^{n,k}_{ab}$ appearing at order $\O(1/|x|^5)$. Also,
$\psi_{ab}\psi^{ab}<0$. Finally, we note that for a neighbourhood
$\mathcal{B}_{a}(i)$ of radius $a$ centred at $i$, one has that
\begin{equation}
|x|^8\psi_{ab}\psi^{ab}\in E^\infty(\mathcal{S}\setminus\{i^\prime\}), \label{eli_condition}
\end{equation}
where $E^\infty(\mathcal{B})=\{f+|x|g \; | \; f,g\in
C^\infty(\mathcal{B})\}$. Similar behaviour is obtained on
$\mathcal{S}\setminus\{i\}$ for $|y|^8\psi_{ab}\psi^{ab}$.

\subsection{General existence results}
The assumption on the conformal flatness of the 3-metric
$\widetilde{h}_{ab}$ and the choice of solutions to the momentum
constraint made in the previous section fulfil the hypothesis of the
existence results obtained in \cite{DaiFri01}. Thus, one has:

\begin{theorem}[Dain \& Friedrich, 2001]
If $\widetilde{h}_{ab}$ is conformally flat, and
$\widetilde{\chi}_{ab}$ is given via equation (\ref{psi_ab}), then
there is a unique, positive solution $\vartheta$ to the Licnerowicz
equation (\ref{cf_hamiltonian}) with the boundary conditions
(\ref{boundary_2}) and (\ref{boundary_4}). Moreover, by virtue of
(\ref{eli_condition}) one has that on
$\mathcal{S}\setminus\{i^\prime\}$:
\begin{equation}
\label{theta_local_p}
\vartheta= \frac{1}{|x|}+ W,
\end{equation}
where $W\in E^\infty(\mathcal{S}\setminus\{i^\prime\})$.
 \end{theorem}

Because of $W\in E^\infty(\mathcal{S}\setminus\{i^\prime\})$, the
function $W$ can be expanded in a neighbourhood of $i$ as:
\begin{equation}
W= \frac{m}{2}+ \big(d_ax^a +m^\prime|x|\big) + \big(q_{ab}x^a x^b + d^\prime_a x^a|x| \big) + \big( o_{abc}x^a x^b x^c + q^\prime_{ab}x^a x^b |x|\big)
+\mathcal{O}(|x|^4),
\end{equation} 
where $m$ is the ADM mass of the initial data, and $m^\prime$, $d_a$
and $d^\prime_a$, $q_{ab}$ and $q^\prime_{ab}$, $o_{abc}$ are,
respectively, monopolar, dipolar, quadrupolar and octupolar terms. In
reference \cite{Val04a} it has been shown that, without loss of
generality, one can make use of the translational freedom in the
setting to remove the dipolar term, $d_a$, in the previous
expansion. Proceeding in such a way, we are left with
\begin{equation}
W=\frac{m}{2}+ m^\prime|x|+ \big(q_{ab}x^a x^b + d^\prime_a x^a|x| \big) + \big( o_{abc}x^a x^b x^c + q^\prime_{ab}x^a x^b |x|\big)
+\mathcal{O}(|x|^4).
\end{equation} 
The explicit calculations carried out in \cite{Val04e} show that in fact $m^\prime=0$ and $d^\prime_{a}=0$.

Alternatively, one could have used the spherical coordinates
associated with the normal coordinates $x^a$. Due to our assumption on
axial symmetry the expansions are given by\footnote{In reference
\cite{Val04e}, on which the present work is based, a certain type of
unitary representations of $SU(2,\Complex)$,
$T_{m\phantom{k}j}^{\phantom{m}k}$, is used. Here,instead, we make use
of the more standard spin-weighted spherical harmonics
${}_sY_{nm}$. The relation between the two sets of functions is given
by
\[
{}_sY_{nm}\mapsto (-\mbox{i})^{s+2n-m}\sqrt{\frac{2n+1}{4\pi}}T_{2n\phantom{n-m}n-s}^{\phantom{2n}n-m}.
\]
Accordingly, we have slightly changed our notation with respect to
that of \cite{Val04e}.}  :
\begin{equation}
W=\frac{m}{2}+ \frac{1}{2!}r^2 \sum_{l=0}^2 
\omega_{2;l} Y_{l0}+ \frac{1}{3!}r^3 \sum_{l=0}^3 
\omega_{3;l} Y_{l0} +\mathcal{O}(r^4), \label{W_spherical}
\end{equation}
where $r=|x|$, and $\omega_{p;l,m}\in \Real$. 

The function $W$ satisfies the elliptic equation:
\begin{equation}
\Delta W =-\frac{1}{8}\psi_{ab}\psi^{ab}(1/|x|+W)^{-7}. \label{Yamabe} 
\end{equation}
Thus, the coefficients $\omega_{2;l,m}$ and $\omega_{3;l,m}$ are
determined by the boundary conditions
(\ref{boundary_1})-(\ref{boundary_4}) and the parameters of
$\psi_{ab}$ ---that is, $\mathcal{A}$, $\mathcal{J}$, $L_{0,n;k}$ and
$L_{4,n;k}$. Thus, sometimes we shall write
$W=W[\mathcal{A},J,L](x)$. The elliptic nature of (\ref{Yamabe})
reveals that $W$ contains information of global nature. The local part
---around infinity--- is encoded in the $1/|x|$ term of
(\ref{theta_local_p}) corresponding to the conformal flatness of the
data.

From equation (\ref{Yamabe}) and
taking into account the boundary conditions
(\ref{boundary_1})-(\ref{boundary_2}) one gets the integral
representation:
\begin{equation}
W=\frac{1}{4\pi}\int_{\Real^3} \frac{\psi_{ab}(x^\prime)\psi^{ab}(x^\prime)}{8(1/|x^\prime|+ W(x^\prime))^7}\frac{1}{|x-x^\prime|} dx^\prime. \label{integral_W}
\end{equation} 
The function $-1/4\pi|x-x^\prime|$ corresponds to the Green's
function of the Laplacian on $\Real^3$. 

For later use, we show that the function $W=W[\mathcal{A},J,L](x)$
depends ---at least--- in a $C^1$ fashion on the parameters
$\mathcal{A}$, $\mathcal{J}$, $L_{0,n;k}$ and $L_{4,n;k}$ which
determine the tensor $\psi_{ab}$.

\begin{lemma}
The solution $W$ of equation (\ref{Yamabe}), with $\psi_{ab}$ given by
equation (\ref{psi_ab}) is $C^1$ in the parameters $\mathcal{A}$,
$\mathcal{J}\in \Real$ and $L_{0,n;k}$, $L_{4,n;k}\in \Complex$
\end{lemma}

To see this, we rewrite equation (\ref{Yamabe}) as 
\begin{equation}
\mathcal{G}(\mathcal{J};W)=\Delta W+\frac{1}{8}\psi_{ab}\psi^{ab}(1/|x|+W)^{-7}=0, \label{implicit}
\end{equation}
and concentrate on the dependence of $W$ with respect to $\mathcal{J}$
---the analysis with respect to the other parameters of $\psi_{ab}$ is
similar. The mapping $\mathcal{G}: \Real\times
W^{2,p}(\mathcal{S})\longrightarrow L^p(\mathcal{S})$, $p\geq 2$,
where $W^{2,p}(\mathcal{S})$ denotes the Sobolev space of functions on
$\mathcal{S}$ with $L^p$-integrable weak second order derivatives, is
at least a $C^2$ mapping (in the $L^p$ norm) between Banach spaces
---actually, it can be shown to be $C^\infty$, but $C^2$ will be
enough for our applications. That $\mathcal{G}\in C^0(\Real\times
W^{2,p},L^p)$ can be shown using
\begin{eqnarray}
&&\hspace{-2cm}\| \mathcal{G}(\mathcal{J},W)-\mathcal{G}(\mathcal{J}_0,W_0)\|_{L^p}\leq \|\Delta W-\Delta W_0\|_{L^p}\nonumber \\
&&\hspace{3cm}+C\|(1+|x|W)^{-7}\|_{L^p}\||x|^7 \psi_{ab}\psi^{ab}(\mathcal{J})-|x|^7\psi_{ab}\psi^{ab}(\mathcal{J}_0)\|_{L^p} \nonumber \\
&&\hspace{3cm} + C\||x|^7\psi_{ab}\psi^{ab}(\mathcal{J}_0)\|_{L^p}\|(1+|x|W)^{-7}-(1+|x|W_0)^{-7}\|_{L^p},
\end{eqnarray}
and noting that $\Delta W$ converges to $\Delta W_0$ is $W$ goes to
$W_0$ as $W^{2,p}$ is a Banach space, and that $\psi_{ab}\psi^{ab}$ is
in our case analytic function of $\mathcal{J}$. To establish the $C^1$
character of the map $\mathcal{G}$ one has to perform a similar
discussion with the G\^ateaux derivative of $\mathcal{G}$,
$\mbox{d}_G\mathcal{G}: \Real^2\times W^{2,p}\times
W^{2,p}\longrightarrow L^{2,p}$, which is given by
\begin{equation}
\mbox{d}_G \mathcal{G}(\mathcal{J},j;W,w)=\Delta w-\frac{1}{8}j\partial_\mathcal{J}\psi_{ab}\psi^{ab}(1/|x|+W)^{-7}+\frac{7}{8}w\psi_{ab}\psi^{ab} (1/|x|+W)^{-8}.
\end{equation} 
Again, the analyticity of $\psi_{ab}$ with respect to
$\mathcal{J}$ is crucial to establish the continuity of
$\mbox{d}_G\mathcal{G}$. Higher order derivatives are dealt with in a
similar fashion.

Now, let us consider a solution $W_0$ to equation (\ref{Yamabe}),
corresponding to a certain value, $\mathcal{J}_0$, of the angular
momentum, so that
\begin{equation}
\mathcal{G}(\mathcal{J}_0,W_0)=0.
\end{equation} 
The existence of such a solution is guaranteed by the theorem 1. The
linearised operator
\begin{equation}
\mathcal{L}w=\Delta w-\frac{7}{8}\psi_{ab}\psi^{ab}(1/|x|+W)^{-8}w,
\end{equation}
$\mathcal{L}:W^{2,p}\longrightarrow L^p$ is an isomorphism: due to
$\psi_{ab}\psi^{ab}>0$ the only solution to $\mathcal{L}w=0$ is $w=0$
(seen using integration by parts); further, $\mathcal{L}$ is
selfadjoint and, thus, using the Fredholm alternative it is also
surjective ---see eg. \cite{DaiFri01}. So, from the implicit function
theorem we know there exists solution $W=W[\mathcal{J}](x)$ to equation
(\ref{implicit}) for $\mathcal{J}$ sufficiently close to
$\mathcal{J}_0$ ---cfr. \cite{AmbPro95}. By unicity, this solution has
to coincide with that one given in theorem 1. More importantly for our
purposes is that $W\in C^2(U,X)$ ---actually $C^\infty$!--- for some
$U\subset \Real$, $\mathcal{J}_0\in U$ and $X\subset W^{2,p}$,
$W(\mathcal{J}_0)\in X$. Poinwise differentiability with respect to
$\mathcal{J}$ follows from the Sobolev embedding theorem: if
$(k-r-\alpha)/n\geq 1/p$ then $W^{k,p}(U)\subset C^{r,\alpha}(U)$,
with $U\subset \Real^n$ ---see, for example \cite{Aub82}. In our case
$k=2$ (at least!), $n=1$, $r\geq 1$ and $p$ suitably large ---say,
$p\geq 2$. Note that if one only had $W\in C^1(U,X)$ then the
pointwise continuity ---with respect to $\mathcal{J}$--- of
$\partial_\mathcal{J}W$ can not be concluded.

\section{Obstructions to the smoothness of null infinity}

As mentioned in the introduction, the calculations carried out in
\cite{Val04e} provide asymptotic expansions for the components of the
Weyl spinor which allow to relate the structure of the initial data
with the radiative properties of the development. Here, and in what
follows we will concentrate our attention on the null infinity
associated with the asymptotic end $i$. What happens at the null
infinity assigned to the second asymptotic end, $i^\prime$, will not
be of relevance for our purposes. Assuming axial symmetry, the
components of the Weyl tensor can be written as:
\begin{equation}
\phi_j=\sum_{p=0} \sum_{l=|2-j|}^p \alpha_{j,p;l}(\tau) r^p\;\; {}_{2-j}Y_{l0}, \label{phi_j}
\end{equation}
where $j=0,\ldots,4$. The coefficients $\alpha_{j,p;l}(\tau)$ for
$p=0,\ldots,4$ are \emph{polynomial} in $\tau$.

For $p=5$, they are polynomial in $\tau$ for $l=0,1,3$. However, for $l=2$ one
has that
\begin{equation}
\alpha_{j,5;2}(\tau)=\Upsilon_{5;2}\left( (1-\tau)^{7-j}\mathcal{P}_j(\tau)\ln(1-\tau) + (1+\tau)^{3+j}\mathcal{P}_{4-j}(\tau)\ln(1+\tau)\right) +\mathcal{Q}(\tau), \label{alpha_52k}
\end{equation}
where $\mathcal{P}_j(\tau)$, $\mathcal{P}_{4-j}(\tau)$ and $\mathcal{Q}(\tau)$ 
are some polynomials in $\tau$. In particular, $\mathcal{P}_j(\pm1)\neq 0$, 
$\mathcal{P}_{4-j}(\pm1)\neq 0$.

Similarly, for $p=6$ the coefficients $\alpha_{j,p;l}(\tau)$ are polynomial
for $p=0,1,4$, while the coefficients of the $l=2$ harmonics are of the form
\begin{equation}
\alpha_{j,6;2}= \Upsilon^+_{6;2}(1-\tau)^{8-j}\ln(1-\tau)\mathcal{P}_j(\tau) + \Upsilon^-_{6;2}(1+\tau)^{4+j}\mathcal{P}_{4-j}(\tau)\ln(1+\tau)+\mathcal{Q}(\tau), \label{alpha_62k}
\end{equation}
where again $\mathcal{P}_j(\tau)$, $\mathcal{P}_{4-j}(\tau)$ and
$\mathcal{Q}(\tau)$ are some polynomials in $\tau$,
$\mathcal{P}_j(\pm1)\neq 0$, $\mathcal{P}_{4-j}(\pm1)\neq 0$
---different to those in equation
(\ref{alpha_52k}). Finally,
\begin{equation}
\alpha_{j,6;3}(\tau)=\Upsilon_{6;3}\left( (1-\tau)^{8-j}\mathcal{P}_{j+1}(\tau)\ln(1-\tau) + (1+\tau)^{4+j}\mathcal{P}_{5-j}(\tau)\ln(1+\tau)\right) +\mathcal{Q}(\tau) \label{alpha_63k}.
\end{equation}
The coefficients $\alpha_{j,p;q}(\tau)$ for $p\geq 7$ are expected to
exhibit a similar pattern. In particular, it is conjectured that there
is an infinite hierarchy of coefficients $\Upsilon^\pm_{p;q}$
associated with logarithmic terms of the form discussed above. These
details will not be relevant for our analysis.

The coefficients $\Upsilon_{5;2}$, $\Upsilon^\pm_{6;2}$ and
$\Upsilon_{6;3}$ ---\emph{the obstructions to the smoothness of null
infinity}--- are given in terms of the freely specifiable data by:
\begin{eqnarray}
&& \Upsilon_{5;2}=9\sqrt{\frac{5}{\pi}}m^2 \omega_{2;2,2}+\frac{37602}{199}m \mathcal{J}^2-\frac{3099}{199}\sqrt{6}m^2 \mbox{i}\alpha + \frac{2046}{199}\sqrt{6}m \mbox{i}\beta_I+\frac{448}{199}\sqrt{6}\mbox{i}\delta_I, \label{Upsilon522}\\ 
&& \Upsilon^\pm_{6;2}=\frac{2198208}{6965}\mathcal{J}^2\pm \frac{62451}{14}\mathcal{A} \mathcal{J}^2 -\frac{62691}{2408}\sqrt{6}\mbox{i}\mathcal{A}\beta_I +\frac{1791}{14}\sqrt{6}\beta_R \pm\frac{263327}{31605}\mbox{i}\beta_I  \nonumber \\
&& \hspace{2cm}-\frac{116}{43}\sqrt{6}\mbox{i}\mathcal{A}\delta_I-\frac{1791}{7}\sqrt{6}\delta_R\mp\frac{183184}{13545}\sqrt{6}\mbox{i}\delta_I, \label{Upsilon622}\\
&& \Upsilon_{6;3}=12\sqrt{\frac{7}{\pi}}\omega_{3;3,3}-\frac{565753248}{82585}\mbox{i}\mathcal{J}^3+\frac{36399}{415}\sqrt{6} \mathcal{J}\beta_I+\frac{3808}{415}\sqrt{6}\mathcal{J}\delta_I-\frac{7272}{415}\sqrt{30}\mbox{i}\gamma. \label{Upsilon633}
\end{eqnarray} 
where in order to render the above expressions more readable we have set
\begin{subequations}
\begin{eqnarray}
L_{0,-2;4}=\mbox{i}\alpha, && L_{4,-2;4}=-\mbox{i}\alpha, \\
L_{0,-1;4}=\beta_R+\mbox{i}\beta_I, && L_{4,-1;4}=\beta_R-\mbox{i}\beta_I, \\
L_{0,-1,6}=\mbox{i}\gamma, &&  L_{4,-1;6}=-\mbox{i}\gamma, \\
L_{0,0;4}=\delta_R+\mbox{i}\delta_I, &&  L_{4,0;4}=\delta_R-\mbox{i}\delta_I,
\end{eqnarray}
\end{subequations}
in accordance with the reality conditions (\ref{reality}) and the regularity conditions (\ref{regularity}). 

The role of the obstructions (\ref{Upsilon522})-(\ref{Upsilon633}) can
be better understood by noting that the conformal factor $\Theta$
---rendering the representation of the region of spacetime near null
and spatial infinity in which the components of the Weyl tensor
(\ref{phi_j}) are calculated--- is of the form,
\begin{equation}
\Theta=\frac{D_a\vartheta D^a\vartheta}{\vartheta^3}(1-\tau^2),
\end{equation}
where $D_a\vartheta D^a\vartheta/\vartheta^3=\O(\rho)$. The locus of
null infinity is given by $\scri^\pm=\{\tau=\pm 1, r\neq 0\}$. In
addition, we define $I=\{\rho=0, |\tau|<1\}$ and $I^\pm=\{\rho=0,
\tau=\pm1\}$ corresponding, respectively, to the \emph{cylinder at
spatial infinity} and the \emph{critical sets} where null infinity
``touches'' spatial infinity. Thus, the coefficients
$\alpha_{j,p;q}(\tau)$ in the expansion (\ref{phi_j}), considered as
functions on $I$ diverge on the critical sets $I^\pm$ if the relevant
obstructions are not satisfied. Because of the hyperbolic nature of
the equations governing the evolution of the gravitational field, one
expects that these singularities will propagate along null infinity.

\subsection{A particular choice of the data} 
Inspired by the above discussion, we are interested in constructing
an initial data such that:
\begin{subequations} 
\begin{eqnarray}
&& \Upsilon_{5;2}=\Upsilon^+_{6;2}=\Upsilon_{6;3}=0, \label{cond_1} \\
&& \Upsilon^-_{6;2}\neq 0. \label{cond_2}
\end{eqnarray}
\end{subequations}
The evolution of such an initial data set will possess ---modulo an
existence proof for the conformal field equations which is valid up to
the critical sets, $I^\pm$, which is not yet available, see
e.g. \cite{Fri03b}--- a null infinity where $\scri^+$ and $\scri^-$
have different smoothness. Assuming that the conditions (\ref{cond_1})
and (\ref{cond_2}) are sharp, a rough count using
(\ref{alpha_52k})-(\ref{alpha_63k}) suggests that the time development
of our initial data will have an $\scri^-$ which, at best, will be of
class $C^3$ and an $\scri^+$ of at least class $C^4$. In particular it
is claimed the following:
\begin{conjecture}
  The time development of an initial data set
  $(\widetilde{h}_{ab},\widetilde{\chi}_{ab})$ such
  that the conditions (\ref{cond_1}) and (\ref{cond_2}) hold admits a
  peeling future null infinity and a polyhomogeneous past null
  infinity. More precisely, using and an adapted gauge the components of
  the Weyl tensor\footnote{The convention used here to denote the
  different components of the Weyl tensor is exactly the opposite of
  the standard Newman-Penrose (NP) convention. That is, our $\phi_0$
  corresponds to NP's $\psi_4$, and so on.} decay, near $\scri^+$, as
\begin{equation}
\phi_0=\O(1/r), \quad \phi_1=\O(1/r^2), \quad \phi_2=\O(1/r^3), \quad \phi_3=\O(1/r^4), \quad \phi_4=\O(1/r^5).
\end{equation}
While near $\scri^-$ the decay will be
\begin{equation}
\phi_4=\O(1/r), \quad \phi_3=\O(1/r^2), \quad \phi_2=\O(1/r^3), \quad \phi_1=\O(1/r^4), \quad \phi_0=\O(\ln r/r^5).
\end{equation} 

\end{conjecture}

That is, the resulting spacetime is expected to peel at $\scri^+$, but
not at $\scri^-$. For a discussion of the notion of polyhomogeneity at
null infinity and non-peeling spacetimes see for example
\cite{ChrMacSin95}.

\section{The main result}
In this section it is shown how the freely specifiable parameters
contained in the initial data discussed in section 2 ---the scalars
$\mathcal{A}$, $\mathcal{J}$ and $\alpha$, $\beta_R$, $\beta_I$,
$\gamma$, $\delta_R$ and $\delta_I$---can be used to construct initial
data satisfying the conditions (\ref{cond_1}) and (\ref{cond_2}). To
this end, one has to find a way of controlling the coefficients
$\omega_{2;2}$ and $\omega_{3;3}$ in the expansion
(\ref{W_spherical}).

\begin{theorem}
Given $\mathcal{A}_0\in\Real$, $\mathcal{A}_0\neq0$ and a conformally
flat initial data set with second fundamental form given by
(\ref{psi_ab}) and satisfying the boundary conditions
(\ref{boundary_2}) and (\ref{boundary_4}), there exists a
neighbourhood $\mho\subset \Real^3$, $(0,0,0)\in\mho$ such that if
$(\mathcal{J},\beta_R,\beta_I)\in\mho$, then there are unique
$\mathcal{A}=\mathcal{A}(\mathcal{J},\beta_R,\beta_I)$,
$\alpha=\alpha(\mathcal{J},\beta_R,\beta_I)$,
$\delta_R=\delta_R(\mathcal{J},\beta_R,\beta_I)$,
$\delta_I=\delta_I(\mathcal{J},\beta_R,\beta_I)$,
$\gamma=\gamma(\mathcal{J},\beta_R,\beta_I)$ with
$\mathcal{A}(0,0,0)=\mathcal{A}_0$,
$\alpha(0,0,0)=\delta_R(0,0,0)=\delta_I(0,0,0)=\gamma(0,0,0)=0$, such
that
\[
\Upsilon_{5;2}=\Upsilon^+_{6;2}=\Upsilon_{6;3}=0,
\]
The condition 
\[
\Upsilon^-_{6;2}\neq 0,
\] 
is satisfied if $\mathcal{J}\neq 0$, $\beta_R\neq0$, $\beta_I\neq0$.
\end{theorem}

Thus, given a particular time asymmetric, conformally flat
Schwarzschild initial data set, there is always a (non-linear)
perturbation of it ---actually an open set of them--- satisfying the
conditions (\ref{cond_1}) and (\ref{cond_2}). Note that these
perturbations have necessarily non-vanishing angular momentum;
furthermore, from a careful analysis of the proof it stems that all the
parameters of the second fundamental form have to be non-zero
---justifying the somewhat complicated-looking choice of the extrinsic
curvature made in 2.2.  The different parts of the proof will be
discussed in the sequel.

\subsection{Integral representations of $\omega_{2;2}$ and $\omega_{3;3}$.}
Our first task is to obtain some expressions which enable us to control
the expansion coefficients in (\ref{W_spherical}). The explicit
calculations of \cite{Val04e} show that:
\begin{equation}
W=\frac{m}{2}+ \frac{1}{2!}r^2 \omega_{2;2} Y_{20}+ \frac{1}{3!}r^3 \sum_{l=0}^3 \omega_{3;l} Y_{lm} +\mathcal{O}(r^4),
\end{equation}
in $\mathcal{S}\setminus\{i^\prime\}$, where
\begin{subequations}
\begin{eqnarray}
&& \omega_{3;0}=-2\sqrt{\pi}\left(\frac{3J^2}{4}+\frac{\mathcal{A}^2}{96}\right), \\
&& \omega_{3;1}=3\sqrt{\frac{\pi}{5}}J^2, \\
&& \omega_{3;2}=0.
\end{eqnarray}
\end{subequations}
On the other hand, the coefficients $\omega_{2;2}$ and $\omega_{3;3}$ do 
not admit such closed and neat expressions. They contain information of global 
nature. Thus, in order to control them we have to resort to the integral 
representation (\ref{integral_W}). 

We begin by noting that:
\begin{equation}
\partial_{rr}W= \omega_{2;2}Y_{20}+\mathcal{O}(r).
\end{equation}
In order to differentiate the integral representation
(\ref{integral_W}) we have to proceed with care, for the Green
function $1/|x-x^\prime|$ is singular if $x=x^\prime$. We begin by
letting
\begin{equation}
F(x)=\frac{\psi_{ab}(x)\psi^{ab}(x)}{(1/|x|+ W(x))^7}.
\end{equation}
Due to the existence results of Theorem 1, $F(x)\in
E^\infty(\mathcal{S}\setminus\{i^\prime\})$, and $F(x)=\O(|x|)$. Accordingly, $\partial_{r}^{(k)}F$, $k=1,2,3,$ are regular at $r=|x|=0$. And so,
\begin{eqnarray}
&& W=\frac{1}{32\pi} \int_{\Real^3} F(x^\prime) \frac{1}{|x-x^\prime|}d^3x^\prime, \nonumber \\
&& \phantom{W}=\frac{1}{32\pi} \int_{\Real^3} F(x-x^\prime) \frac{1}{|x^\prime|}d^3x^\prime.
\end{eqnarray}
Frome where
\begin{equation}
\partial^{(k)}_rW=\frac{1}{32\pi} \int_{\Real^3} \partial^{(k)}_r F(x-x^\prime) \frac{1}{|x^\prime|}d^3x^\prime,
\end{equation}
with $k=1,2,3$. Consequently, we find that
\begin{equation}
\omega_{2;2}=\frac{1}{32\pi}\int_{\Real^3} \langle Y_{20}| \partial_{rr}F(-x^\prime)\rangle \frac{1}{|x^\prime|}d^3 x^\prime, \label{omega_220}
\end{equation}
where
\begin{equation}
\langle Y_{lm} | H(\theta,\varphi) \rangle =\int_0^{2\pi} \int_0^\pi \overline{Y}_{lm}(\theta,\varphi) H(\theta,\varphi) \sin \theta d\theta d\varphi,
\end{equation}
denotes the component of $H$ on the $(l,m)$-harmonic. Similarly, one finds that
\begin{equation}
\omega_{3;3}=\frac{1}{32\pi}\int_{\Real^3} \langle Y_{30}| \partial_{rrr}F(-x^\prime)\rangle \frac{1}{|x^\prime|}d^3 x^\prime. \label{omega_330}
\end{equation}

\subsection{Satisfying the conditions for a ``smoothness asymmetric null infinity''.}
In order to see how the conditions (\ref{cond_1}) and (\ref{cond_2})
can be satisfied, we begin by considering imaginary parts of the
obstructions
(\ref{Upsilon522})-(\ref{Upsilon633}):
\begin{subequations}
\begin{eqnarray}
&& \mbox{Im}(\Upsilon_{5;2})=-\frac{3099}{199}\sqrt{6}m^2 \alpha + \frac{2046}{199}\sqrt{6}m \beta_I+\frac{448}{199}\sqrt{6}\delta_I=0, \label{im_upsilon520}\\
&& \mbox{Im}(\Upsilon^+_{6;2})= -\frac{62691}{2408}\sqrt{6}\mathcal{A}\beta_I +\frac{263327}{31605}\beta_I -\frac{116}{43}\sqrt{6}\mathcal{A}\delta_I-\frac{183184}{13545}\sqrt{6}\delta_I=0, \label{im_upsilon_620} \\
&& \mbox{Im}(\Upsilon_{6;3})=-\frac{565753248}{82585}\mathcal{J}^3-\frac{7272}{415}\sqrt{30}\gamma=0. \label{im_upsilon_630}
\end{eqnarray}
\end{subequations} 
The condition (\ref{im_upsilon520}) can be used to solve for $\alpha$
in terms of $\beta_I$ and $\delta_I$. Condition (\ref{im_upsilon_620})
can be used to solve for $\delta_I$ in terms of $\mathcal{A}$ and
$\beta_I$. Finally, condition (\ref{im_upsilon_630}) can be used to
write $\gamma$ in terms of $\mathcal{J}$, rendering $\gamma\propto
\mathcal{J}^3$. On similar lines
\begin{equation}
\mbox{Re}(\Upsilon^+_{6;2})=\frac{2198208}{6965}\mathcal{J}^2+\frac{62451}{14}\mathcal{A} \mathcal{J}^2  +\frac{1791}{14}\sqrt{6}\beta_R-\frac{1791}{7}\sqrt{6}\delta_R=0,
\end{equation}
can be used to solve for $\delta_R$ in terms of $\mathcal{A}$,
$\mathcal{J}$ and $\beta_R$. Hence, we are left with only 4 free real
parameters: $\mathcal{A}$, $\mathcal{J}$, $\beta_R$, $\beta_I$.

Satisfying $\mbox{Re}(\Upsilon_{5;2})=0$ and
$\mbox{Re}(\Upsilon_{6;3})=0$ is much more subtle. For this we turn
to the integral representations (\ref{omega_220}) and
(\ref{omega_330}) of the coefficients $\omega_{2;2}$ and
$\omega_{3;3}$ of $W$. Consequently, one finds that
\begin{subequations} 
\begin{eqnarray}
&&\hspace{-2.2cm}\mbox{Re}(\Upsilon_{5;2})=\frac{9}{32\pi}\sqrt{\frac{5}{\pi}}m^2\int_{\Real^3} \langle Y_{20}|\partial_{rr}F(-x^\prime)\rangle \frac{1}{|x^\prime|} d^3x^\prime + \frac{37602}{199}m\mathcal{J}^2=0, \label{integral_1}\\
&&\hspace{-2.2cm} \mbox{Re}(\Upsilon_{6;3})=\frac{3}{8\pi}\sqrt{\frac{7}{\pi}}\int_{\Real^3} \langle Y_{30}|\partial_{rrr}F(-x^\prime)\rangle\frac{1}{|x^\prime|} d^3x^\prime +\frac{36399}{415}\sqrt{6} \mathcal{J}\beta_I+\frac{3808}{415}\sqrt{6}\mathcal{J}\delta_I=0. \label{integral_2}
\end{eqnarray}
\end{subequations}
Now, we note that $W=W[\mathcal{A},\mathcal{J},\beta_R,\beta_I](x)$.
That is, the function $W$ depends on the the choice of the free
parameters $\mathcal{A}$, $\mathcal{J}$, $\beta_R$, $\beta_I$. For
example, given our choice of boundary conditions, if
$\mathcal{A}=\mathcal{J}=\beta_R=\beta_I=0$, then the initial set
reduces to that of the standard time symmetric, conformally flat slice
of the Schwarzschild spacetime, and accordingly $W=m/2$. More
crucially for our purposes, if $\mathcal{A}=\mathcal{A}_0\neq0$, but
we keep $\mathcal{J}=\beta_R=\beta_I=0$, then the initial data
corresponds to a conformally flat, time asymmetric slice in the
Schwarzschild spacetime, and ---see
\cite{Val04e}---
\begin{equation}
W=\frac{m}{2}-\frac{1}{576}\mathcal{A}^2|x|^2+\O(|x|^4). \label{W_A}
\end{equation}
The latter fact suggests the use of an ``implicit function
theorem''-type of argument to see if there are some values of
$\mathcal{A}$, $\mathcal{J}$, $\beta_R$ and $\beta_I$ such that
(\ref{integral_1}) and (\ref{integral_2}) hold.

Given that the choice $\mathcal{A}=\mathcal{A}_0\neq0$,
$\mathcal{J}=\beta_R=\beta_I=0$ corresponds to Schwarzschild data, we
have that
\begin{subequations}
\begin{eqnarray}
&& \mbox{Re}(\Upsilon_{5;2})[\mathcal{A}_0,0,0,0]=0,  \label{u5_0}\\
&& \mbox{Re}(\Upsilon_{6;3})[\mathcal{A}_0,0,0,0]=0;  \label{u6_0}
\end{eqnarray}
\end{subequations}
the smoothness of of static spacetimes at the critical sets where null
 infinity ``touches'' spatial infinity has been discussed in
 \cite{Fri04}. On the other hand, using lemma 1,
 $\mbox{Re}(\Upsilon_{5;2})$ and $\mbox{Re}(\Upsilon_{6;3})$
 depend in, at least, $C^1$ fashion on $\mathcal{A}$. From (\ref{psipsi}) and
 (\ref{W_A}) it follows that
\begin{subequations}
\begin{eqnarray}
&& \partial_\mathcal{A} \mbox{Re}(\Upsilon_{5;2})[\mathcal{A}_0,0,0,0]\neq0, \\
&& \partial_\mathcal{A} \mbox{Re}(\Upsilon_{6;3})[\mathcal{A}_0,0,0,0]\neq0.
\end{eqnarray}
\end{subequations}
Thus, if we set
\begin{equation}
\mathcal{F}(\mathcal{A},\mathcal{J},\beta_R,\beta_I)= \bigg(\mbox{Re}(\Upsilon_{5;2})[\mathcal{A},\mathcal{J},\beta_R,\beta_I]
\bigg)^2 + \bigg(\mbox{Re}(\Upsilon_{6;3})[\mathcal{A},\mathcal{J},\beta_R,\beta_I]\bigg)^2, 
\end{equation}
one has that
\begin{subequations}
\begin{eqnarray}
&& \mathcal{F}(\mathcal{A}_0,0,0,0)=0, \\
&& \partial_\mathcal{A}\mathcal{F}(\mathcal{A}_0,0,0,0)\neq 0,
\end{eqnarray}
\end{subequations}
so that from the implicit function theorem ---see
e.g. \cite{AmbPro95,CouJoh89}--- there exists a neighbourhood $\mho
\subset \Real^3$, $(0,0,0)\in \mho$ and a $C^1$ function
$\mathcal{A}(\mathcal{J},\beta_R,\beta_I)$,
$\mathcal{A}(0,0,0)=\mathcal{A}_0$
such that
\begin{equation}
\mathcal{F}(\mathcal{A}(\mathcal{J},\beta_R,\beta_I), \mathcal{J},\beta_R,\beta_I)=0, \quad (\mathcal{J},\beta_R,\beta_I)\in\mho.
\end{equation}
The latter proves essentially proves our main result. A careful
inspection reveals that $\Upsilon^-_{6;2}\neq 0$ if $J\neq0$,
$\beta_R\neq0$ and $\beta_I\neq0$.  We note that a similar argument
with the choice $\mathcal{A}=\mathcal{J}=\beta_R=\beta_I=0$ does not
work for it implies
$\partial_\mathcal{A}\psi_{ab}\psi^{ab}=\partial_\mathcal{A} W=0$ so
that $\partial_\mathcal{A}\mathcal{F}(0,0,0,0)=0$ ---the implicit
function theorem can not be applied. That is, the Schwarzschild
initial data that one is perturbing have to be time necessarily asymmetric.

\section{Conclusions}
The initial data we have constructed in this article is a non-linear
perturbation of time asymmetric Schwarzschild initial data. It is
conformally flat and has non-vanishing angular momentum. Its time
development should give rise to a radiative spacetime which could be
interpreted as a distorted Schwarzschild black hole. The crucial 
remaining problem is to obtain an existence proof for the evolution
equations which is valid up to (and beyond) the critical sets where
null infinity ``touches'' spatial infinity ---see the discussions in
\cite{Fri03b,Fri04}. Obtaining such a proof will require deeper
insights into the structure of spatial infinity than the ones
currently available.

\section*{Acknowledgements}
I thank H. Friedrich and S. Dain for helpful discussions and CM
Losert-Valiente Kroon for a careful reading of the
manuscript. Valuable criticism and suggestions from an anonymous
referee are thankfully acknowledged.  This project was started during
a visit to the Max Planck Institute for Gravitational Physics (Albert
Einstein Institute) in Golm, Germany, and continued during a stay at
the Isaac Newton Institute in Cambridge as a part of the programme on
``Global Problems in Mathematical Relativity''. I thank these
institutions for their hospitality. This research is funded by an
EPSRC Advanced Research Fellowship.


\begin{thebibliography}{10}

\bibitem{AmbPro95}
A.~Ambrosetti \& G.~Prodi,
\newblock {\em A primer of nonlinear analysis},
\newblock Cambridge University Press, 1995.

\bibitem{Aub82}
T.~Aubin, \newblock {\em Nonlinear Analysis on
Manifolds. Monge-Amp\`ere Equations.} Springer, 1982.

\bibitem{BeiOMu96}
R.~Beig \& N.~O. Murchadha,
\newblock {\em The momentum constraints of General Relativity and spatial
  conformal isometries},
\newblock Comm. Math. Phys. {\bf 176}, 723 (1996).

\bibitem{BeiOMu98}
R.~Beig \& N.~O. Murchadha,
\newblock {\em Late time behavior of the maximal slicing of the Schwarzschild black hole},
\newblock Phys. Rev. D {\bf 57}, 4728 (1998).

\bibitem{ChrDel03}
P.~T. Chru\'{s}ciel \& E.~Delay,
\newblock {\em On mapping properties of the general relativistic constraint
  operator in weighted function spaces, with applications},
\newblock in {\tt gr-qc/0301073}.

\bibitem{ChrDel02}
P.~T. Chru\'{s}ciel \& E.~Delay,
\newblock {\em Existence of non-rivial, vacuum, asymptotically simple
  spacetimes},
\newblock Class. Quantum Grav. {\bf 19}, L71 (2002).

\bibitem{ChrMacSin95}
P.~T. Chru\'{s}ciel, M.~A.~H. Mac{Callum}, \& D.~B. Singleton,
\newblock {\em Gravitational waves in general relativity {X}{I}{V}. {Bondi}
  expansions and the ``polyhomogeneity'' of $\scri$},
\newblock Phil. Trans. Roy. Soc. Lond. A {\bf 350}, 113 (1995).

\bibitem{Cor00}
J.~Corvino,
\newblock {\em Scalar curvature deformations and a gluing construction for the
  {Einstein} constraint equations},
\newblock Comm. Math. Phys. {\bf 214}, 137 (2000).

\bibitem{CorSch03}
J.~Corvino \& R.~Schoen,
\newblock {\em On the asymptotics for the Einstein Constraint Vacuum
  Equations},
\newblock At gr-qc/0301071.

\bibitem{CouJoh89}
R.~Courant \& F.~John,
\newblock {\em Introduction to Calculus and Analysis II/1},
\newblock Springer, 1989.

\bibitem{Dai04a}
S.~Dain,
\newblock {\em Asymptotically flat and regular Cauchy data},
\newblock in {\em The conformal structure of space-time. Geometry, analysis,
  numerics}, edited by J.~Frauendiener \& H.~Friedrich, page 161, Springer,
  2003.

\bibitem{DaiFri01}
S.~Dain \& H.~Friedrich,
\newblock {\em Asymptotically flat initial data with prescribed regularity at
  infinity},
\newblock Comm. Math. Phys. {\bf 222}, 569 (2001).

\bibitem{EstWahlChrDeWSmaTsi73}
F.~Estabrook, H.~Wahlquist, S.~Christensen, B.~DeWitt, L.~Smarr, \& E.~Tsiang,
\newblock {\em Maximally slicing a black hole},
\newblock Phys. Rev. D {\bf 7}, 2814 (1973).

\bibitem{Eva98}
L.~C.~Evans, \newblock{\em Partial Differential Equations}, American Mathematical Society, 1998.

\bibitem{Fri98a}
H.~Friedrich,
\newblock {\em Gravitational fields near space-like and null infinity},
\newblock J. Geom. Phys. {\bf 24}, 83 (1998).

\bibitem{Fri03a}
H.~Friedrich,
\newblock {\em Conformal Einstein evolution},
\newblock in {\em The conformal structure of spacetime: Geometry, Analysis,
  Numerics}, edited by J.~Frauendiener \& H.~Friedrich, Lecture Notes in
  Physics, page~1, Springer, 2002.

\bibitem{Fri03b}
H.~Friedrich,
\newblock {\em Spin-2 fields on Minkowski space near space-like and null
  infinity},
\newblock Class. Quantum Grav. {\bf 20}, 101 (2003).

\bibitem{Fri04}
H.~Friedrich,
\newblock {\em Smoothness at null infinity and the structure of initial data},
\newblock in {\em 50 years of the Cauchy problem in general relativity}, edited
  by P.~T. Chru\'{s}ciel \& H.~Friedrich, Birkhausser, 2004.

\bibitem{GilTru98}
D.~Gilbarg \& N.~S.~Trudinger, 
\newblock {\em Partial Differential Equations of Second Order},
Springer, 1998.

\bibitem{Pen63}
R.~Penrose,
\newblock {\em Asymptotic properties of fields and space-times},
\newblock Phys. Rev. Lett. {\bf 10}, 66 (1963).

\bibitem{Pen02}
R.~Penrose,
\newblock {\em Twistor geometry of conformal infinity},
\newblock in {\em The conformal structure of spacetime: Geometry, Analysis,
  Numerics}, edited by J.~Frauendiener \& H.~Friedrich, Lecture Notes in
  Physics, page~1, Springer, 2002.

\bibitem{PenRin86}
R.~Penrose \& W.~Rindler,
\newblock {\em Spinors and space-time. {V}olume 2. {S}pinor and twistor methods
  in space-time geometry},
\newblock Cambridge University Press, 1986.

\bibitem{Rei73}
B.~L. Reinhart,
\newblock {\em Maximal foliations of extended Schwarzschild space},
\newblock J. Math. Phys. {\bf 14}, 719 (1973).

\bibitem{Som80}
P.~Sommers,
\newblock {\em Space spinors},
\newblock J. Math. Phys. {\bf 21}, 2567 (1980).

\bibitem{Val04d}
J.~A. Valiente~Kroon,
\newblock {\em Does asymptotic simplicity allow for radiation near spatial
  infinity?},
\newblock Comm. Math. Phys. {\bf 251} (2004).

\bibitem{Val04a}
J.~A. Valiente~Kroon,
\newblock {\em A new class of obstructions to the smoothness of null infinity},
\newblock Comm. Math. Phys. {\bf 244}, 133 (2004).

\bibitem{Val04e}
J.~A. Valiente~Kroon,
\newblock {\em Time asymmetric spacetimes near null and spatial infinity. I.
  Expansions of developments of conformally flat data},
\newblock Class. Quantum Grav. {\bf 23}, 5457 (2004).

\bibitem{Val05a}
J.~A. Valiente~Kroon,
\newblock {\em Time asymmetric spacetimes near null and spatial infinity. II.
  Expansions of developments of initial data sets with non-smooth conformal
  metrics},
\newblock Class. Quantum Grav. {\bf 22}, 1683 (2005).

\bibitem{WalWil79a}
M.~Walker \& C.~M. Will,
\newblock {\em Relativistic {Kepler} problem. {I}. Behavior in the distant past
  of orbits with gravitational radiation damping},
\newblock Phys. Rev. D {\bf 19}, 3483 (1979).

\bibitem{WalWil79b}
M.~Walker \& C.~M. Will,
\newblock {\em Relativistic {Kepler} problem. {I}{I}. Asymptotic behavior of
  the field in the infinite past},
\newblock Phys. Rev. D {\bf 19}, 3495 (1979).

\end{thebibliography}
\end{document}